\documentclass{elsart5p}

\usepackage{graphicx}
\usepackage{bm}
\usepackage[latin1]{inputenc}
\usepackage{amssymb}
\usepackage{amsmath}

\newcommand{\nn}{\nonumber}
\newcommand{\so}{\mathrm{so}}
\newcommand{\uu}[1]{\vc{#1}}
\newcommand{\vc}[1]{\bm{#1}}
\newcommand{\vco}[1]{\hat{\bm{#1}}}

\newcommand{\be}{\begin{eqnarray}}
\newcommand{\ee}{\end{eqnarray}}
\newcommand{\bea}{\begin{eqnarray}}
\newcommand{\eea}{\end{eqnarray}}

\begin{document}

\begin{frontmatter}

\title{Spin dephasing due to a random Berry phase}

\author[tfp]{Pablo San-Jose},
\author[tfp]{Gerd Sch\"on},
\author[tfp]{Alexander Shnirman}, and
\author[Budapest,tfp]{Gergely Zarand}

\address[tfp]{Institut f\"{u}r Theoretische
Festk\"{o}rperphysik and DFG-Center
    for Functional Nanostructures (CFN), Universit\"{a}t Karlsruhe,
    D-76128 Karlsruhe, Germany.}
\address[Budapest]{Institute of Physics,
Technical University Budapest, Budapest, H-1521, Hungary.}

\begin{abstract}
We investigate relaxation and dephasing of an electron spin
confined in a semiconductor quantum dot and subject to
spin-orbit coupling. Even in vanishing magnetic field, $\vc
B = 0$, slow noise coupling to the electron's orbital
degree of freedom leads to dephasing of the spin due to a
random, in general non-Abelian Berry phase acquired by the
spin. For illustration we first present a simple
quasiclassical description, then consider a model with 2
orbital states only, and finally present a perturbative
quantum treatment appropriate for an electron in a
realistic (roughly parabolic, not too strongly confining)
quantum dot. We further compare the effect of different
sources of noise. While at large magnetic fields phonons
dominate the relaxation processes, at low fields
electron-hole excitations and possibly $1/f$ noise may
dominate.
\end{abstract}

\end{frontmatter}

\section{Introduction \label{sec:introduction}}

The demonstrations of coherent single-electron spin control
and measurement~\cite{Elzerman04,Petta05,Koppens06} in
semiconductor quantum dots have opened exciting
perspectives for solid state quantum information processing
with spin qubits \cite{Loss98}. More recent
work~\cite{Kato05,Hankiewicz06,Engel05,Awschalom07} has
revealed further potential of spin coherence, which greatly
extends the possibilities of next-generation spintronic
devices. The key behind these emerging technologies is the
long spin coherence time in semiconductor materials. Spins,
unlike orbital electron degrees of freedom, do not couple
directly to the various sources of electric noise present
in typical solid-state environments.

Most of the traditional techniques for addressing and
manipulating spins in semiconductors have revolved around
some form of electron spin resonance (ESR), be it through
external magnetic fields~\cite{Koppens06} or effective
internal ac fields
based on the spin-orbit interaction. Indeed, spin-orbit
interaction has been proposed theoretically as a way of
coherently controlling the spin of confined electrons
purely by electrical
means~\cite{Duckheim06,Bulaev06,Golovach06,Stano06b,Tang06},
and important experimental progress has been made in this
direction \cite{Stotz04,Kato04}. By the same token, it has
long been understood~\cite{Abrahams57,Dyakonov72} that
spin-orbit interaction is one of the main mechanisms by
which electron spins decay and lose coherence in
semiconductor heterostructures
\cite{Khaetskii00,Khaetskii01,Woods02,Golovach04}.

As we will discuss in this paper, in the particular case of
an electron confined in a quantum dot, a time-dependent
(fluctuating or controlled) electric field introduces via
the spin-orbit coupling a non-Abelian geometric phase (a
generalization of Berry phases) into the spin evolution.
This connection between spin-orbit interaction and
geometric phases has been noted previously in the context
of perturbative analysis of the spin decay of trapped
electrons~\cite{sanjose06}. A similar connection had been
discussed for free electrons in the presence of disorder
scattering~\cite{Serebrennikov04}.

The geometric character of spin evolution under electric
fields has striking consequences both for spin-orbit
mediated spin relaxation and decoherence as well as for
coherent spin manipulation strategies. Geometric spin
evolution under controlled gating is potentially robust,
since it is not affected by gate timing errors and certain
control voltage inaccuracies. In the case of spin decay,
the non-Abelian character of the spin precession under a
noisy electric environment results in a saturation of spin
relaxation rates at low magnetic fields~\cite{Abrahams57}
through a fourth order (in the spin-bath coupling) process
previously overlooked in the literature
\cite{Khaetskii01,Stano06,Semenov06}. This spin decay
mechanism, which can be called {\it geometric dephasing},
requires two independent noise sources coupled to two
non-commuting components of the electron spin, whereby the
non-Abelian properties of the SU(2) group become relevant.
\footnote{A different phenomenon, also called geometric
dephasing, was discussed in Ref.~\cite{Whitney05}. There
geometric manipulations of spins in finite magnetic fields
and the presence of dissipation were considered and
path-dependent (geometric) contributions to dephasing were
found.} To second order in the spin-bath coupling we also
note that a different source of fluctuations other that
piezoelectric phonons, namely electron-hole excitations in
the metallic environment (ohmic fluctuations), dominate the
spin relaxation at low magnetic fields. The reason is the
higher density of ohmic fluctuations at low energies as
compared to phonons.

The present paper is organized as follows. In
Sec.~\ref{sec:concepts} we will present qualitatively the
main concepts and consequences of the geometric character
of the electrically induced spin precession in leading
order in the ratio between dot size to spin-orbit length,
$x_0/l_\so$. In Sec.~\ref{sec:toy} we consider a model
system based on only two orbital states of an electron with
spin. This model helps to understand the geometrical
evolution of the spin. In Sec.~\ref{sec:full} we will
perturbatively derive the effective Hamiltonian for an
electron in a quantum dot under a fluctuating electric
field taking into account all orbital states. This will
allow us to analyze the spin relaxation and dephasing under
realistic conditions.

\section{Geometric spin precession of a strongly
confined electron \label{sec:concepts}}

Electric fields applied to a quantum dot structure induce
displacements (and possibly deformations) in the confining
potential. In the presence of spin-orbit interaction this
will lead to a peculiar geometric evolution of the spin
state of the confined electron with important consequences
for the relaxation and manipulation of the spin. We will
approach the problem by first considering the spin
precession due to a geometric phase acquired by the spin of
a strongly confined electron, when it is adiabatically
transported along a {\sl given trajectory } in a 2DEG in
the presence of spin-orbit coupling.

In semiconductor 2D heterostructures the spin-orbit
coupling takes the form ($\hbar=1$ throughout this work)
\begin{eqnarray}
\mathcal{H}_\so&=&\alpha\left(\hat{p}_y\hat{\sigma}_x
-\hat{p}_x\hat{\sigma}_y\right)+\beta\left(\hat{p}_y\hat{\sigma}_y
-\hat{p}_x\hat{\sigma}_x\right)
\nonumber\\
&=& \frac{1}{m}\vco{p}\uu{\lambda}^{-1}_\so\vco{\sigma} \,
. \label{Hsodef}
\end{eqnarray}
Here $\vco{\sigma}/2$ and $\vco{p}$ are the spin and
momentum operators, while $\alpha$ and $\beta$ are the
Rashba and linear Dresselhaus couplings, which can be
lumped into the spin-orbit tensor
\begin{eqnarray}
\uu{\lambda}_\so^{-1}\equiv m\left(\begin{array}{cc} -\beta&-\alpha\\
\alpha&\beta\end{array}\right).
\end{eqnarray}
It sets the scale for the spin-orbit length
$l_\so\equiv\sqrt{|\det\uu\lambda_\so|}=
\left(m\sqrt{|\alpha^2-\beta^2|}\right)^{-1}$. The
effective strength of the spin-orbit effects in a quantum
dot of size $x_0$ is in general proportional to some power
of the ratio $x_0/l_\so$. In typical GaAs/AlGaAs
semiconductor heterostructures $l_\so\sim 1-5\mathrm{\mu
m}$, while $x_0\sim 30-100 \mathrm{nm}$ so that this ratio
is usually quite small, of the order of $0.02$. Other
materials, such as InAs, have a much stronger spin-orbit
length, in the $l_\so\sim 100\mathrm{nm}$ range.

By classical intuition we can anticipate the main effect.
We consider an electron in a very strong confinement,
forced to move along a path $\mathcal{C}$ with trajectory
$\vc R_\mathcal{C}(t)$. Eq. (\ref{Hsodef}) suggests that
the spin-orbit coupling makes the spin precess under an
effective magnetic field $\vc B_\so=\frac{1}{m}\vco
p\uu\lambda_\so^{-1}$, which couples to the spin similar to
a Zeeman term except that the field depends on the
electron's momentum. It raises the question as to what
'value' one should use for operator $\vco p$. For a
strongly confining potential it turns out that we can
simply substitute 
$\vco p \rightarrow m{\dot {\vc R}}_{\mathcal{C}}$. Hence
\begin{eqnarray}
\vc B_\so={\dot {\vc R}}_\mathcal{C}\uu\lambda_\so^{-1}\label{classic}
\end{eqnarray}
From this we derive a spin precession governed by the
following SU(2) operator
\begin{eqnarray}
U_\mathrm{ad}(t)&=&T\exp\left(-i\int_0^tdt\,
\vc B_\so\cdot\vco\sigma\right)\nn\\
&=&P\exp\left(-i\int_\mathcal{C} d\vc R_\mathcal{C}\,
\uu\lambda_\so^{-1}\vco\sigma\right)\label{Uadclassic}
\end{eqnarray}
Here $T$ and $P$ stand for time- and path-ordering
operators, respectively. The label `adiabatic' in
$U_\mathrm{ad}$ refers to the constraint of slow paths,
$|{\dot{\vc{R}}}_\mathcal{C}|\ll x_0\omega_0$, typically
assumed in most works on Berry phases \cite{Berry84}. As is
apparent from Eq. (\ref{Uadclassic}), due the peculiar
dependence of $\vc B_\so$ on the velocity 
${\dot {\vc R}}_\mathcal{C}$, the total ``geometric spin precession" for
propagation along a given path $\mathcal{C}$ depends only
on the geometry of $\mathcal{C}$ itself, not on the time
dependence of ${\vc R}_\mathcal{C}$.

Another line of arguments leading to this result was
pointed out in Ref. \cite{Coish06}. It is based on the
observation that $H_\so$ can be diagonalized to first order
in $x_0/l_\so$ by a canonical transformation
$\exp{\left(-i\vco r\uu\lambda_\so^{-1}\vco
\sigma\right)}$, which in turn implies that in a small dot
the effect of spin-orbit coupling moving along a given path
can be gauged away by a path-dependent gauge transformation
$U_\mathrm{ad}$ that rotates the spin just as in Eq.
(\ref{Uadclassic}).

\begin{figure}
\includegraphics[width=8.5 cm]{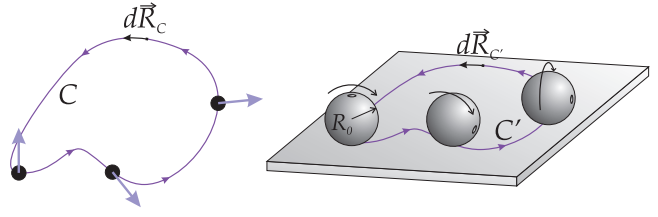}
\caption{The geometric precession due to spin orbit
interaction for an electron adiabatically transported along
a path in a 2DEG (left) is equivalent to the changing
orientation of a sphere rolling on a plane along a path
(right) which is related by a simple transformation to the
electron's path.\label{fig:rolling}}
\end{figure}

The evolution operator $U_\mathrm{ad}$ is a  group element
in SU(2). However, it can also be mapped onto a SO(3)
rotation of a 3D solid, since both groups are isomorphic up
to a sign. The natural question arises, what is the 3D
rotation corresponding to $U_\mathrm{ad}$ for a given path?
Is there an intuitive visualization that tells us how the
spin is rotated as the containing quantum dot is moved?
Remarkably, the SO(3) isomorphic form of $U_\mathrm{ad}$
(changing the SU(2) generators $\vco\sigma/2$ by the SO(3)
equivalent $\vco A$) has a very similar form to the
operator that gives the orientation of a sphere of radius
$R_0$ that rolls on a plane without slipping or spinning
along a path $\mathcal{C'}$ parametrized by
$\vc{R}_\mathcal{C'}(t)$,
\begin{eqnarray}
U_\mathrm{sph}&=&P\exp\left(-i\int_{C'}d\vc
R_\mathcal{C'}\, \uu\lambda_\mathrm{sph}^{-1}\vco A\right)\\
\uu\lambda_\mathrm{sph}^{-1}&=&\frac{1}{R_0}\left(\begin{array}{cc}
0&1\\-1&0\end{array}\right)
\end{eqnarray}
This picture of the geometric spin-precession in terms of a
rolling sphere is illustrated in Fig. \ref{fig:rolling}.
The radius of the sphere is fixed by the spin-orbit length,
$R_0=l_\so/2$. If only Rashba coupling is present there is
nothing more to this mapping, since in such case
$\uu\lambda_\mathrm{sph}\propto\uu\lambda_\so$. However, if
Dresselhaus coupling is also present the paths $C$ and $C'$
are not exactly the same, but are related by a simple 2D
rotation and scaling transformation
$\vc{R}_\mathcal{C'}(t)\uu\lambda_\mathrm{sph}^{-1}
=2\vc{R}_\mathcal{C}(t)\uu{\lambda}_\so^{-1}$.

\begin{figure}
\includegraphics[width=6 cm]{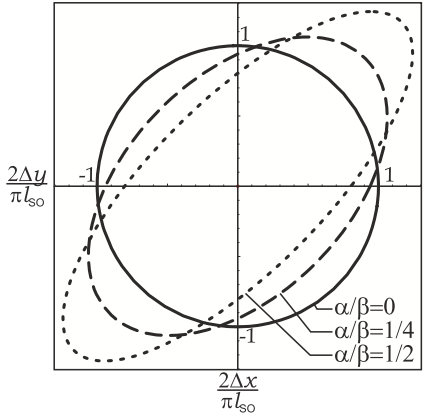}
\caption{The length of a straight path transport of a small
dot required to perform a pseudospin flip as a function of
the angle of the path. The three ellipses correspond to
different ratios $\beta/\alpha$ but the same value of
$l_\so$. Dotted ellipse: $\beta/\alpha = 1/2$; dashed
ellipse: $\beta/\alpha = 1/4$, solid ellipse:
$\beta/\alpha=0$. The area of the ellipses remains constant
and equal to $\pi l_\so^2$.\label{fig:anisotropy}}
\end{figure}

This geometrical picture makes it clear that transporting
the spin along a sufficiently long  straight path
$\mathcal{C}$ in a certain direction in the 2DEG will flip
the pseudospin. In the case of the sphere moving along a
straight path $\mathcal{C'}$ the distance needed for a
specific rotation does not depend on the direction, but for
a spin rotation, due to the nontrivial relation between
$\mathcal{C}$ and $\mathcal{C'}$ when both Rashba and
Dresselhaus couplings are present, the required distance is
non-isotropic in a 2DEG. In the direction $[110]$ the
spin-flip distance can be strongly enhanced. This
non-isotropic property is represented in the
Fig.~\ref{fig:anisotropy} for different ratios of
$\beta/\alpha$. In average, the distance that must be
covered to induce a pseudospin flip is $\pi l_\so/2$, but
the variation as a function of the direction can be large
whenever $\alpha$ approaches $\beta$.

\section{Two orbital states with spin}
\label{sec:toy}

\subsection{The model}

Before analyzing the dynamics of an electron spin in a
typical quantum dot we consider a model system with merely
two orbital levels, $|0\rangle$ and $|1\rangle$, which are
occupied by a single electron with spin-1/2. This model is
not valid to describe electrons confined in (near)
parabolic quantum dots, where all orbital level are
important. Yet, it allows us to get a feeling for the
geometric phases acquired by the pseudo-spin states.

The Hilbert space of the model system is spanned by the
four states $|0,\uparrow\rangle$, $|0,\downarrow\rangle$,
$|1,\uparrow\rangle$, $|1,\downarrow\rangle$. We introduce
two sets of Pauli matrices: $\hat\tau_\alpha$ for orbital
degrees of freedom and $\hat\sigma_\alpha$ for the spin.
Thus, for example, $ \hat\tau_x |0\rangle = |1\rangle $,
while $ \hat\sigma_x |\uparrow\rangle = |\downarrow\rangle
$. In the presence of spin-orbit interaction the
Hamiltonian of the model system reads
\begin{eqnarray}
H_{\rm dot} = -\frac{1}{2}\,\epsilon \hat\tau_z -
\frac{1}{2}\,\hat\tau_y\,\vc{b} \cdot \vco{\sigma}\ ,
\end{eqnarray}
where $\epsilon$ is the orbital level splitting, while the
vector $\vc{b}$ characterizes the Rashba and Dresselhaus
spin-orbit coupling. Choosing real orbital wave functions
one can easily check that this is the most general form of
the spin-orbit coupling allowed by time-reversal symmetry.

Next we introduce the noise which is assumed to couple to
the orbital degree of freedom. The general form of coupling
is
\begin{eqnarray}
H_{\rm int} = -\frac{1}{2}\,\hat\tau_z\,\hat Z -
\frac{1}{2}\,\tau_x\, \hat X\ ,
\end{eqnarray}
where in general $\hat Z$ and $\hat X$ are quantum fields
of a bath governed by the Hamiltonian $H_{\rm bath}$.

For slow noise, i.e. for noise with power spectrum with
much weight at low frequencies, instead of considering the
coupled quantum dynamics of the dot and the bath, it is
sufficient to treat $\hat Z$ and $\hat X$ as classical
stochastic fields, independent of each other. Thus, for
transparency, we shall consider the dynamics of the dot
subject to the external random fields $Z(t)$ and $X(t)$
and, then, average over realizations of these fields.

\begin{figure}
\includegraphics[width=5.5 cm]{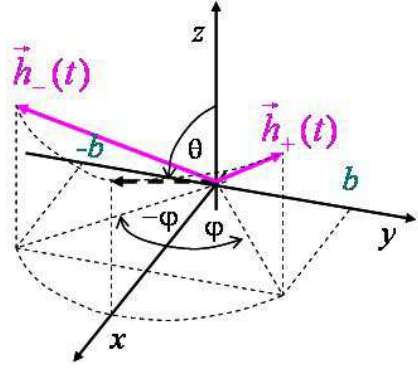}
\caption{\label{fig:magfields} Pseudo-magnetic fields
$\vc{h}_{\pm}$.}
\end{figure}

The resulting total Hamiltonian
\begin{eqnarray}
H_{\rm dot} = -\frac{1}{2}\,\epsilon \hat\tau_z -
\frac{1}{2}\,\hat\tau_y\,\vc{b} \cdot
\vco{\sigma}-\frac{1}{2}\,Z(t)\,\hat\tau_z -
\frac{1}{2}\,X(t)\,\hat\tau_x \ .
\end{eqnarray}
shows that the direction of the vector $\vc{b}$ is the
natural quantization axis for the spin in our problem.
Introducing the corresponding spin basis states
$|+\rangle_{\vc{b}}$ and $|-\rangle_{\vc{b}}$ we see that
the problem factorizes into two subspaces: $|\Psi_{\rm
orbital}\rangle|+\rangle_{\vc{b}}$ and $|\Psi_{\rm
orbital}\rangle|-\rangle_{\vc{b}}$. Within each of the two
subspaces $(\pm)$ the dynamics is governed by the
Hamiltonian
\begin{eqnarray}
H_{\pm}&=& -\frac{1}{2}\,\epsilon \hat\tau_z \mp
\frac{1}{2}\,b\,\hat\tau_y - \frac{1}{2}\,Z(t)\,\hat\tau_z
- \frac{1}{2}\,X(t)\,\hat\tau_x\nonumber\\ &=&
-\frac{1}{2}\,\vc{h}_{\pm}(t) \vco{\tau} ,
\end{eqnarray}
where $b\equiv |\vc{b}|$. We have introduced the
pseudo-magnetic fields $\vc{h}_{\pm}$ acting on the orbital
``pseudo-spin". These fields are illustrated in
Fig.~\ref{fig:magfields}. The fields $\vc{h}_{\pm}$ differ
only in their $y$-component, $h_{\pm,y}=\pm b$. Therefore
we obtain $|\vc{h}_{+}|=|\vc{h}_{-}|$ and -- consistent
with Kramers' theorem -- we find that the energies of the
ground states in the two subspaces,
$|g_+\rangle|+\rangle_{\vc{b}}$ and
$|g_-\rangle|-\rangle_{\vc{b}}$, coincide,
$E_{g,+}=E_{g,-}= -(1/2)|\vc{h}_{+}|$.
\begin{figure}
\includegraphics[width=5.5 cm]{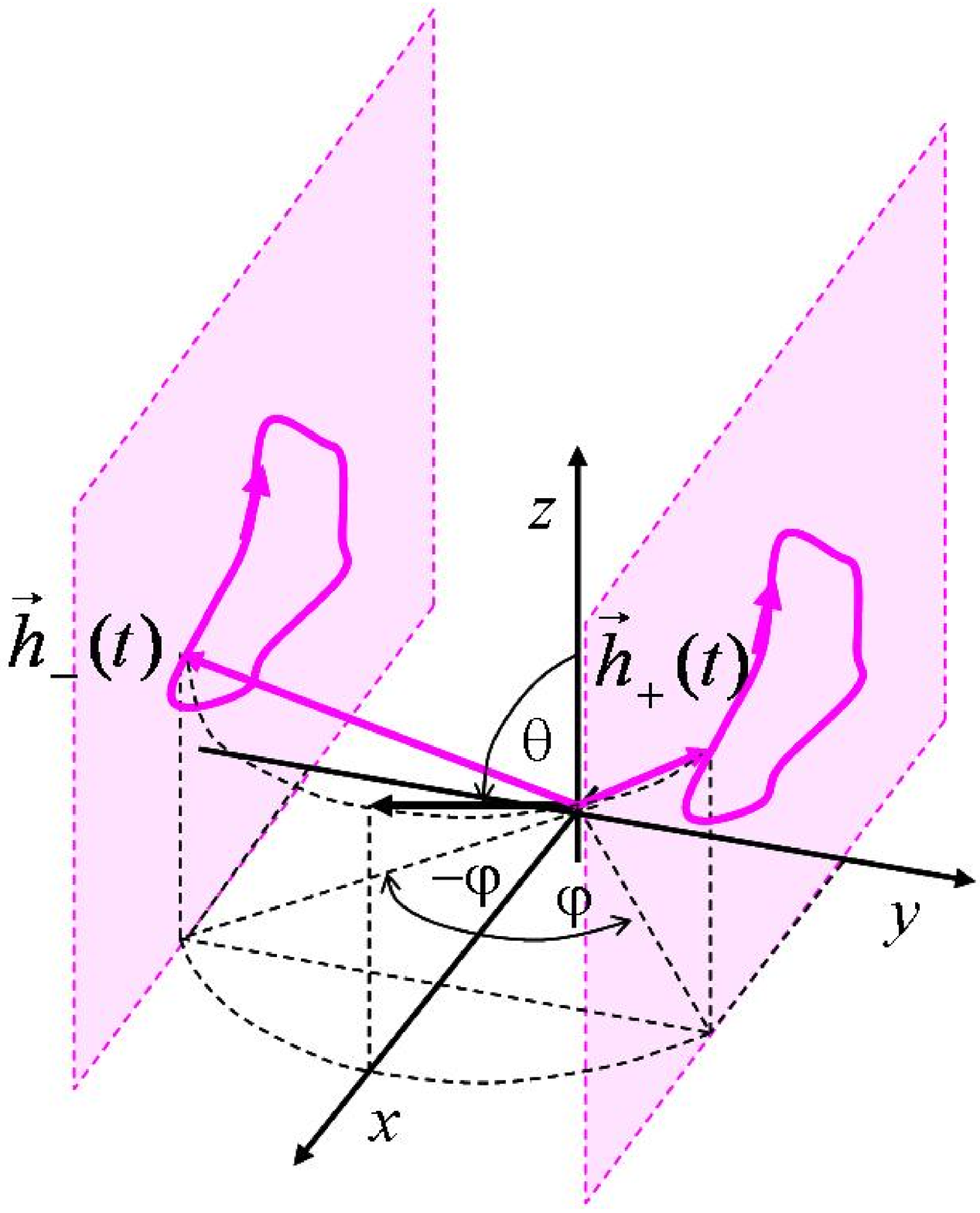}
\caption{\label{fig:contours} Closed contours traversed by
the pseudo-magnetic fields $\vc{h}_{\pm}$.}
\end{figure}

\subsection{Random Berry phase in fluctuating fields}
Next we study what happens when the fields $Z(t)$ and
$X(t)$ vary in time. In particular, we assume that they
traverse slowly a closed contour in the $X-Z$ plane, as
illustrated in Fig.~\ref{fig:contours}. Then the
pseudo-spin orbital eigenstates states follow their
respective pseudo-fields. In addition to the dynamical
phase each state acquires a Berry phase. As one can see
from Fig.~\ref{fig:contours}, the Berry phases acquired by
the two ground states from the subspaces "+" and "-" are
opposite in sign
\begin{eqnarray}
\Phi_{\pm} = \pm \frac{1}{2}\int d\varphi \cos\theta\ ,
\end{eqnarray}
where the angles $\varphi$ and $\theta$ are introduced in
Figs.~\ref{fig:magfields} and \ref{fig:contours}. Since the
dynamical phases for the two ground states are the same,
the relative phase acquired between the two is
\begin{eqnarray}
\Delta\Phi = \Phi_+ - \Phi_- = \int d\varphi \cos\theta\ .
\end{eqnarray}

If we express the angles $\varphi$ and $\theta$ in terms of
the fields $X$ and $Z$ and expand assuming $X,Z \ll
\epsilon,b$, we find after some algebra
\begin{eqnarray}
\Delta\Phi = \int dt
\,\frac{b}{(\epsilon^2+b^2)^{3/2}}\,Z(t)\dot X(t) + ...\ .
\end{eqnarray}
The dots in this relation denote contributions which vanish
for closed contours and, as can be shown, do not cause
dephasing.

Several notes are in order. Since the evolution of the
pseudospin is described here by a single phase
$\Delta\Phi$, rotations due to two different loops in the
$X-Z$ plane commute. Thus we obtain an Abelian Berry phase.
Indeed, as we have seen, the subspaces $+$ and $-$ do not
mix. If, initially, we had a superposition $\alpha
|g_+\rangle|+\rangle_{\vc{b}}+ \beta
|g_-\rangle|-\rangle_{\vc{b}}$, the absolute values
$|\alpha|$ and $|\beta|$ are conserved and only the
relative phase changes due to the adiabatic evolution.

Treating the quantity $Z(t) \dot X(t)$ as a Gaussian
stochastic field we obtain the dephasing rate as
\cite{Makhlin02}
\begin{eqnarray}
\label{toy_dephasing} \Gamma_\varphi =
\frac{1}{2}\,\frac{b^2}{(\epsilon^2+b^2)^3}\,S_{Z\dot
X}(\omega=0)\ ,
\end{eqnarray}
where $S_{Z \dot X}$ is the spectral density of $Z \dot X$.
We estimate this quantity as
\begin{eqnarray}
S_{Z\dot X}(\omega=0) \sim 2\int\limits_0^T d\omega
\,\omega^2 S_X(\omega) S_Z(\omega)\ ,
\end{eqnarray}
where the limitation of the integration by the temperature
can be justified by noting that any field can be treated
classically only at frequencies lower than temperature. A
fully quantum mechanical analysis confirms this assertion.

\section{Electron spin in a quantum dot}
\label{sec:full}

\subsection{Effective Hamiltonian}

Lifting the restriction to two orbital states, we now
consider the full problem of a single electron confined to
a lateral quantum dot by the potential
$V(\hat{\mathbf{r}})$ in the presence of a magnetic field
$\vec B$. To be specific, we assume the field to be
oriented parallel to the plane of the dot, but our
procedure can be generalized to arbitrary directions
\cite{Pablo2}. The static part of the Hamiltonian then
reads
\begin{eqnarray}
H_{\rm 0} =\frac{\mathbf{p}^2}{2m^*}+V(\mathbf{r})-
\frac{g\mu_B}{2}\vec{B}\cdot\vec{\sigma}+H_\so \, .
\end{eqnarray}
The magnetic field couples to the electron through a Zeeman
term with material specific $g$-factor~\footnote{For GaAs
$m^*\approx 0.067\, m_e$, $g\approx -0.44$, longitudinal
and transverse sound velocities $v_l= 4.73\cdot 10^3 m/s$,
$v_t=3.35\cdot 10^3 m/s$, piezoelectric const.\
$h_{14}=1.4\cdot 10^9 eV/m$, density $\rho=5.3\cdot 10^3
Kg/m^3$, $\lambda_{ph}=(e^2h_{14}^2/105(2\pi)^2\rho
(3/v_l^5+4/v_t^5)$. For the quantitative analysis we
considered Dresselhaus coupling with
$\lambda_{SO}=1/(m^*\beta)=1\mu m$.}. The last term,
defined in Eq. \ref{Hsodef}, describes the Dresselhaus
($\beta$) and Rashba spin-orbit couplings ($\alpha$)
between the spin $\vec \sigma$ of the electron and its
momentum \cite{Winkler03}. For a dot of size $x_0$ the
typical energy of the spin-orbit coupling scales as $\sim
\max{\alpha,\beta}/x_0$, while the level spacing scales as
$\omega_0 \sim 1/(m^* x_0^2)$. Therefore, for dots with
small level spacing, $\omega_0 < 1 \;{\rm K}$, the
spin-orbit coupling cannot be treated perturbatively.

Next we account for time-dependent fluctuations of the
electromagnetic field, which add a term
\begin{eqnarray}
\delta V (t) =  X^\mu(t) \; \hat{O}_\mu \label{V(t)}
\end{eqnarray}
to the Hamiltonian. (A summation over repeated indices,
such as $\mu$, is assumed throughout.) The  terms
$\hat{O}_\mu$ denote independent operators in the Hilbert
space of the confined electron (e.g., $x$, $y$,
$x^2,\dots$), while the terms $X^\mu(t)$ denote the
corresponding fluctuating (in general quantum) fields
(e.g., $\delta E_x$, $\delta E_y$, $\nabla_x \delta E_x$,
...). They may be generated by various environments, such
as phonons, localized defects, or electron-hole
excitations. Information about their specific properties is
contained in the spectral functions, to be specified later.
Note that this formulation covers also quadrupolar
fluctuations.

\begin{figure}
\includegraphics[width=8.2 cm]{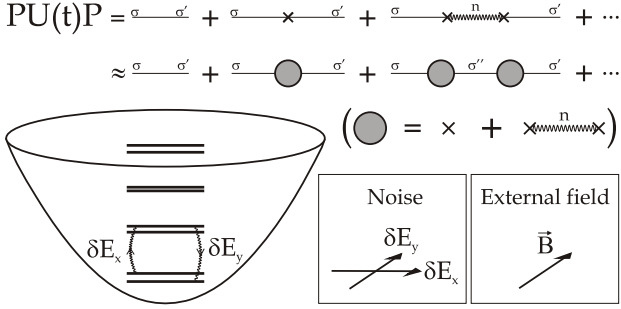}
\caption{ Left: Single electron lateral quantum dot in a
magnetic field, which lifts the ground state degeneracy.
Virtual transitions to excited states are induced by weak
fluctuations of the external fields $\delta E_x(t)$,
$\delta E_y(t)$. Right: Graphical representation of the
evolution operator. Virtual transitions to excited states
$n$ (wavy lines) are integrated out to yield an effective
Hamiltonian within the doublet subspace.
}\label{fig:perturb}
\end{figure}

In case of time-reversal symmetry the ground state of the
dot is two-fold degenerate. This degeneracy is split in an
external magnetic field. If $g \mu_B B \ll \omega_0$, and
as long as the noise is {\it adiabatic} with respect to the
orbital level splitting, $T\ll \omega_0$, the dynamics of
the spin remains constrained to these two states. Under
these conditions (following the method described in
Ref.~\cite{Hutter06})
 we can derive an effective Hamiltonian for the
two lowest eigenstates $|\sigma= \pm\rangle$ of $H_0$ by
expanding the evolution operator $U(t) = \mathrm{T\; exp}\{
-i\int_0^t dt' \;{\delta V}_{\rm int}(t')\}$ and projecting
to the subspace $\{|\sigma\rangle\}$ as sketched in
Fig.~\ref{fig:perturb}. This results in
\begin{eqnarray}
&&PU(t)P  =   1-i\int_0^tdt_1P \delta V_{\rm int}(t_1) P\nn \\
&&-\int \int_{t_1> t_2}  P \delta V_{\rm int}(t_1) P \delta
V_{\rm int}(t_2)P
\label{eq:expansion} \\
&&-\int \int_{t_1> t_2}  P \delta V_{\rm int}(t_1) (1-P)
\delta V_{\rm int}(t_2)P +\ldots\; \nn .
\end{eqnarray}
Here $\delta V_{\rm int}(t)$ denotes the fluctuating part
of the Hamiltonian in the interaction representation and $P
= \sum_\sigma |\sigma\rangle\langle \sigma|$. We separated
terms that involve direct transitions between the two
lowest states from transitions via excited states. In the
spirit of an adiabatic approximation, these latter
processes can be integrated out to yield an effective
Hamiltonian in the two-dimensional subspace. Technically,
this is performed by introducing  slow and fast variables,
$t \equiv (t_1 + t_2)/2$ and  $\tau \equiv t_1 - t_2$, in
the last term of Eq.~(\ref{eq:expansion}),
\begin{eqnarray}
\sim \;  e^{-i t(\epsilon_\sigma-\epsilon_{\sigma'}) -i\tau
\left[\frac{1}{2}(\epsilon_\sigma+\epsilon_{\sigma'})-\epsilon_n\right]}
\; \delta V_{\sigma n}(t_1) \delta V_{n\sigma'} (t_2)\;,
\nonumber
\end{eqnarray}
expanding the interaction potential in $\tau $ as $ \delta
V(t_{1,2}) \approx \delta V(t) \pm
\frac{\tau}{2}\;\frac{d}{dt}{\delta  V}(t) +\dots$, and
integrating with respect to $\tau$. Here $\epsilon_\sigma$
and $\epsilon_n$ denote the eigenenergies of the lowest
doublet and higher eigenstates of $H_0$, respectively. In
this way the last term in Eq.~(\ref{eq:expansion}) becomes
{\em local in time}. Retaining only processes up to 2nd
order, we find an effective Hamiltonian within the
lowest-energy two-dimensional subspace, characterized by
the `pseudospin' Pauli matrices  $\tilde\sigma_{x,y,z}$,
\begin{eqnarray}
H_{\mathrm{eff}}& =& -\frac{1}{2}B_{\rm eff}\;
{\tilde\sigma_z} + X^\mu
\vec{C}^{(1)}_\mu\cdot\vec{\tilde\sigma} + X^\mu
X^\nu\vec{C}^{(2)}_{\mu\nu}\cdot\vec{\tilde\sigma} \nn
\\
&+&\frac{1}{2}\left(\dot{X}^\mu
X^\nu-X^\mu\dot{X}^\nu\right)\vec{C}^{(3)}_{\mu\nu}\cdot\vec{\tilde\sigma}
\; . \label{Heff}
\end{eqnarray}
Due to the spin-orbit coupling, which is not assumed to be
weak, eigenstates do not factorize into orbital and spin
sectors (hence the term `pseudospin'). The static effective
field, ${\vec B}_{\rm eff} \equiv (\epsilon_{+} -
\epsilon_{-})\hat{z}$, accounts for the spin-orbit
renormalization of the $g$-factor and defines the $\hat{z}$
direction in the doublet space. The couplings $\vec
C^{(i)}$, determining the effective fluctuating magnetic
fields felt by the pseudospin, are given by
\begin{eqnarray}
&&\left[\vec{C}^{(1)}_\mu\cdot\vec{\tilde\sigma}\right]_{\sigma,\sigma'}=\hat{O}^\mu_{\sigma
\sigma'}\ ,\\
&&\left[\vec{C}^{(2)}_{\mu\nu}\cdot\vec{\tilde\sigma}\right]_{\sigma,\sigma'}=-
{\sum_n}^\prime \frac{\hat{O}^\mu_{\sigma n}\hat{O}^\nu_{n
\sigma'}}
{\frac{\epsilon_\sigma+\epsilon_{\sigma'}}2-\epsilon_n}\ ,\\
&&\left[\vec{C}^{(3)}_{\mu\nu}\cdot\vec{\tilde\sigma}\right]_{\sigma,\sigma'}=
-i\,{\sum_n}^\prime \frac{\hat{O}^\mu_{\sigma
n}\hat{O}^\nu_{n \sigma'} }
{\left(\frac{\epsilon_\sigma+\epsilon_{\sigma'}}{2}-\epsilon_n\right)^2}\
.
\end{eqnarray}
The summation is restricted to excited states of higher
doublets $n\ne \sigma,\sigma'$. We do not provide explicit
expressions for the eigenenergies $\epsilon_{\sigma}$,
$\epsilon_{n}$, matrix elements $\hat{O}^\mu_{\sigma
\sigma'}$, $\hat{O}^\mu_{\sigma n}$, or couplings $\vec
C^{(i)}$, but below we will evaluate them numerically and
provide quantitative estimates for a generic model. We
further note that both $\vec{C}^{(1)}_\mu$ and
$\vec{C}^{(3)}_{\mu\nu}$ turn out to be transversal to
${\vec B}_{\rm eff}$, therefore contributing only to
relaxation, whereas $\vec{C}^{(2)}_{\mu\nu}$ has in general
also a parallel component that leads to pure dephasing.

\subsection{Geometric phases in $B=0$ case}

In time-reversal symmetric situation, (i.e. for $B=0$), the
first three terms of Eq.~(\ref{Heff}) vanish identically
\cite{Khaetskii01,Pablo2}. Only the last term survives, and
leads to spin dephasing. It has a {\em geometrical} origin.
To demonstrate this, let us assume that the fluctuating
(adiabatic) fields $X_\mu$ are classical. We introduce the
instantaneous ground states of the Hamiltonian, $|\Phi_n
(t)\rangle \equiv |\Phi_n(X_\mu(t))\rangle$ defined through
the equation \be [H_0+\delta V(X_\mu)] |\Phi_n(X_\mu)
\rangle = E_n(X_\mu) |\Phi_n(X_\mu) \rangle\;. \ee Noting
that, to lowest order perturbation theory, the two
degenerate instantaneous ground states are simply  given by
$ |\Phi_\sigma(X_\mu) \rangle \approx |\sigma\rangle +
{\sum_n}^\prime |n\rangle {\langle n|\delta
V|\sigma\rangle} /({\epsilon_\sigma-\epsilon_n}) \;,$ we
can rewrite the last term in Eq.~(\ref{Heff}) in the
familiar form
\be H^{\mathrm{eff}}_{\sigma \sigma'}(B = 0) = -i \langle {\frac{d
\Phi_\sigma}{dt}  | \Phi_{\sigma'} }\rangle \;,
\label{eq:berry}
\ee which shows clearly that the last term is due to a generalized
(possibly non-Abelian) Berry
phase~\cite{Berry84,Wilczek84,Mead87} acquired in a
degenerate 2D subspace. In vanishing magnetic field,
Eq.~(\ref{eq:berry}) can be shown to hold to all orders of
perturbation theory within the adiabatic
approximation~\cite{Pablo2}. If at least two linearly
independent fluctuating fields couple to the dot, they can
produce a random Berry phase for the system and cause
geometric dephasing at $B=0$. When more noise components
are present, the Berry phase may become non-Abelian and all
components of the spin may decay.

\subsection{Relaxation and dephasing times}

So far, our treatment has been rather general, applicable
for arbitrary noise properties and dot geometries. In its
full glory, Eq.~(\ref{Heff}) describes the motion of the
pseudo-spin coupled to three fluctuating ``magnetic
fields". In general, the dynamics induced by these
non-commuting fields is complicated. To obtain a
qualitative understanding of the dynamics we analyze the
spin relaxation and pure dephasing times~\cite{Bloch57},
$T_1$ and $T_2^*$ (with $1/T_2 = 1/2T_1+1/T_2^*$). They are
defined only for sufficiently strong effective fields,
$B_{\rm eff}\gg 1/T_1, \;1/T^{*}_2$. In the limit $B=0$ we
evaluate what we call the geometrical dephasing time
$T_{\rm geom}$. For the quantitative estimate we consider a
parabolic confining potential, $V({\bf r}) = \frac{m^*
\omega_0^2}{2} |{\bf r}|^2$ with level spacing $\omega_0$
and typical size $x_0=1/\sqrt{\omega_0 m^*}$. Furthermore,
we take  into account only dipolar fluctuations, $\hat X
\equiv e \delta E_x x_0$ and $\hat Y \equiv e \delta E_y
x_0$ coupling to the operators ${\hat O}_X \equiv x/x_0$
and ${\hat O}_Y \equiv y/x_0$, respectively. We assume the
two components $\hat X $ and $\hat Y$ to be independent of
each other, but to have identical noise spectra,
$S_X(\omega) = S_Y(\omega) = S(\omega) = \pi
\varrho(\omega) {\rm coth}(\omega/2T)$, with
$\varrho(\omega)$ being the spectral function of the
bosonic environment (phonons or photons).

The spectral function $\varrho(\omega)$ for phonons can be
estimated along the lines of Ref.~\cite{Khaetskii01}. For
the parameters specified in Ref.~\cite{Golovach04} we find
for piezoelectric phonons in typical GaAs heterostructures
at low frequencies, $\varrho_{\rm ph} (\omega) = x_0^2 \;
\lambda_{\rm ph} \; \omega^3$ with $\lambda_{\rm ph}
\approx 4 \cdot 10^{-6} {\rm K}^{-2} {\rm nm}^{-2}$. With
these parameters we obtain relaxation rates generated by
the first term in Eq.~(\ref{Heff}) that coincide with those
of Ref.~\cite{Golovach04} at not too high values of the
field~\footnote{For larger frequencies the approximation
$\varrho_{\rm ph} (\omega) = x_0^2 \; \lambda_{\Omega} \;
\omega^3$ is not valid since the wavelength of relevant
phonons becomes comparable to the size of the dot.}.
Similar values are obtained for the parameters of
Ref.~\cite{Stano05}. For Ohmic fluctuations the spectral
function is linear at low frequencies, $\varrho_\Omega
(\omega) = \lambda_\Omega \;\omega$
\cite{Leggett87,Weiss99}. The prefactor $\lambda_\Omega$
depends on the dimensionless impedance of the circuit,
$\lambda_\Omega\sim \frac { e^2} h \;{\rm Re }[Z]$. For
typical values of the sheet resistance of the 2-DEG
($10^2-10^3\Omega/\Box$) we estimate it to be in the range
$0.1 > \lambda_\Omega > 0.01$. For $1/f$ noise the power
spectrum is $S(\omega) = \lambda_{1/f} /|\omega|$. We will
further comment on its strength below.

We first estimate the contributions $T_1^{(i)}$ and
$T_2^{*(i)}$, derived from the three terms ($i=1,2,3$) in
Eq.~(\ref{Heff}), for a non-vanishing in-plane magnetic
field, $B_{\rm eff}\gg 1/T_1, \;1/T^{*}_2$. The coupling
${\vec C}^{(1)}$ turns out to be
 perpendicular to ${\vec B}_{\rm eff}$ \cite{Golovach04}, and
for low magnetic fields and weak spin-orbit coupling
 is proportional to $|{\vec C}^{(1)}|\sim
\frac{B}{x_0 \omega_0^2} \;\max\{{\alpha}, {\beta}\}$. This
fluctuating field therefore contributes to the $T_1$ energy
relaxation only, \be \frac{1}{T_1^{(1)}} = 2 \left(
|\vec{C}^{(1)}_{X}|^2 + |\vec{C}^{(1)}_{Y}|^2\right)
\;S_X(B_{\rm eff})\ . \ee It scales as $1/T_1^{(1)}\sim B^2
\max \{B,T \}$ for Ohmic dissipation and as $\sim  B^4
\max\{B,T\}$ for phonons. As a consequence, for dots with
level spacing in the range $\omega_0 \approx 1 \dots 10
\;{\rm K}$ Ohmic fluctuations dominate over phonons for low
fields with  $B < 1 \dots 3 \;{\rm T}$.

The second term, $|{\vec C}^{(2)}|\sim
\frac{B}{x_0^2\omega_0^4} \max\{\alpha ^2 , \beta ^2\}$
gives rise to both relaxation and dephasing.~\footnote{Note
that there is no discrepancy with the scaling
$\max\{\alpha, \beta\}$ quoted in Ref.~\cite{Khaetskii01},
since they include higher (multipolar) spin-flipping phonon
contributions neglected here.} The two rates are \bea
\frac{1}{T_1^{(2)}} = 4\left(|\vec{C}^{(2,\perp)}_{XX,s}|^2
+ |\vec{C}^{(2,\perp)}_{YY,s}|^2 +2
|\vec{C}^{(2,\perp)}_{XY,s}|^2\right) S_{XY}(B_{\rm
eff})\;, \nonumber
\\
\frac{1}{T_2^{*(2)}} =
4\left(|\vec{C}^{(2,\parallel)}_{XX,s}|^2 +
|\vec{C}^{(2,\parallel)}_{YY,s}|^2 +2
|\vec{C}^{(2,\parallel)}_{XY,s}|^2\right) S_{XY}(0)\;,
\nonumber
\\
S_{XY}(\omega) = \frac{\pi}{2} \int d\tilde \omega\;
\frac{\varrho(\frac{ \omega + \tilde\omega}{2})
\varrho(\frac{\omega - \tilde\omega}{2})} {1- \cosh(\tilde
\omega/2T)/\cosh(\omega/2T) }\ , \nonumber \eea with
$\vec{C}^{(2,\perp/\parallel)}_{\mu\nu,s}$ denoting the
symmetrized component of ${\vec C}^{(2)}_{\mu\nu}$
perpendicular/parallel to ${\vec B}_{\rm eff}$. Thus, for
Ohmic dissipation ${1}/{T_1^{(2)}}$ vanishes as $\sim  B^2
\max \{B^3,T^3 \}$, while for phonons it scales as $\sim
B^2 \max \{B^7,T^7 \}$.

$\vec{C}^{(3)}$ is also perpendicular to ${\vec B}_{\rm
eff}$. Its contribution to the relaxation is \bea
\frac{1}{T_1^{(3)}} & = & 2 |\vec{C}^{(3)}_{XY,a}|^2
S_{\dot X Y - X\dot Y }(B_{\rm eff})\;,\\
\nonumber
 S_{\dot X Y - X\dot Y }(\omega) &=&\frac{\pi}{2} \int d\tilde \omega\;
\frac{\tilde\omega^2\varrho(\frac{ \omega +
\tilde\omega}{2}) \varrho(\frac{\omega - \tilde\omega}{2})}
{1- \cosh(\tilde \omega/2T)/\cosh(\omega/2T) } \; , \eea
with $\vec{C}^{(3)}_{\mu\nu,a}$ being the anti-symmetrized
component of ${\vec C}^{(3)}_{\mu\nu}$. Most importantly,
with $|{\vec C}^{(3)}_{XY}|\sim \frac{1}{x_0^2\omega_0^4}
\max\{\alpha ^2 ,\beta ^2\}$, the rate ${1}/{T_1^{(3)}}$
approaches a non-zero value at low fields, $1/T_1,1/T_2^*
\ll B\ll T$, and scales as $\sim \max \{B^5,T^5 \}$ for
Ohmic dissipation and $\sim \max \{B^9,T^9 \}$ for phonons.

Finally,  at $B=0$ the geometric dephasing rate is given by
${1}/{T_1^{(3)}}$, extrapolated to zero field \bea
\frac{1}{T_{\rm geom}} =  2 |\vec{C}^{(3)}_{XY,a}|^2
S_{\dot X Y - X\dot Y }(\omega\sim 0)\ . \eea In our
example with only two noise components this process
dephases only the components of the spin perpendicular to
$\vec{C}^{(3)}_{XY}$.

\begin{figure}
\includegraphics[width=8.6 cm]{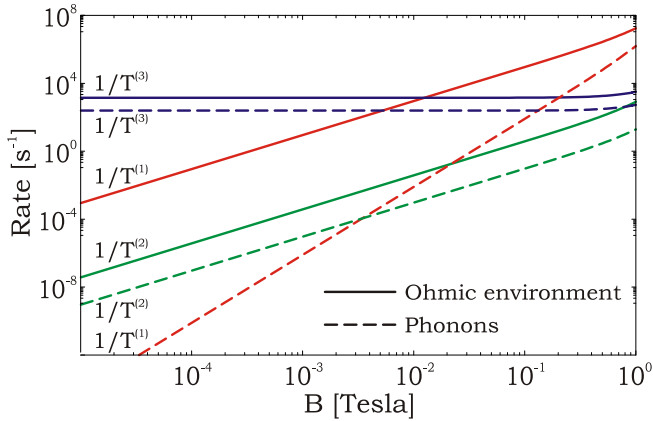}
\caption{\label{fig:rates} Spin relaxation rates  for a
GaAs quantum dot with level spacing $\omega_0 = 1{\rm K}$
as function of the Zeeman field. We chose a temperature $T=
100 \;{\rm mK}$ and Ohmic coupling $\lambda_\Omega = 0.05$.
Below $B^*\approx 1\; {\rm T}$ spin relaxation is dominated
by coupling to Ohmic fluctuations. For $B < B^{**}\approx
15 \;{\rm mT}$ geometrical spin relaxation due to coupling
to Ohmic fluctuations dominates. For all  $B$ values
plotted, the Bloch-Redfield consistency requirement,
$B_{\mathrm{eff}}\gg 1/T^{(i)}$, is satisfied.}
\end{figure}

The relaxation rates corresponding to the different terms
in Eq.~(\ref{Heff}) and various noise sources  are shown in
Fig.~\ref{fig:rates}.
Clearly, for external fields $B< B^* \approx 1  \dots
3\;{\rm T}$ Ohmic fluctuations provide the leading
relaxation mechanism. The crossover field $B^*$ is not very
sensitive to the specific value of the spectral parameter
$\lambda_\Omega$ and is independent of the spin-orbit
coupling. Below a second crossover field, $B^{**} \approx
15 {\rm mT}$, the geometric dephasing induced by Ohmic
fluctuations starts to dominate. This second crossover
scale is very sensitive to the spin-orbit coupling and
temperature, scaling as $B^{**}\sim \max\{\alpha,\beta\}
(1/x_0) (T/\omega_0)^2$. E.g.\ for a level spacing
$\omega_0\sim 1{\rm K}$ and temperature $T=100 \;{\rm mK}$
($T=50 \;{\rm mK}$) the Berry phase mechanism gives a
relaxation time of the order of $700\;{\rm \mu s}$
($20\;{\rm ms}$). For even lower temperatures or  smaller
dots with level spacing $\omega_0\sim 10 {\rm K}$ the $B\to
0$ relaxation time is quickly pushed up to the range of
seconds.

Finally, we comment on the effect of $1/f$ noise. In most
cases, the non-symmetrized correlators for $1/f$ noise,
needed to calculate correlators as $S_{XY}$ or $S_{XX}$,
are not known. Yet, for $|\omega|\ll T$ we can provide an
estimate $S_{XY}(\omega) \approx
\int\limits_{-T}^{T}\frac{d\tilde\omega}{2\pi} S_X(\omega -
\tilde\omega) S_Y(\tilde\omega)$, and similarly for
$S_{\dot{X}Y-X\dot{Y}}$. The $B=0$ geometrical dephasing
rate due to the $1/f$ noise can be estimated as $T_{\rm
geom}^{-1}\approx|{\vec C}^{(3)}_{XY}|^2
\lambda_{1/f}^2(T)\, \omega_c$, where $\omega_c$ is the
upper frequency cut-off for the $1/f$
noise~\cite{Ithier05}. Accounting for the high-frequency
(Ohmic) noise, sometimes observed to be associated with the
$1/f$
noise~\cite{Astafiev04,Shnirman05,Faoro05,Astafiev06}, the
estimate becomes $T_{\rm geom}^{-1}\approx|{\vec
C}^{(3)}_{XY}|^2 \lambda_{1/f}^2(T)\, T$. While the $1/f$
noise of background charge fluctuations is well studied in
mesoscopic systems, the amplitude of the $1/f$ noise of the
electric field in quantum dot systems is yet to be
determined. If we assume that this noise is due to
two-level systems at the interfaces of the top gate
electrodes, we conclude that in the parameter range
explored here, the effect of $1/f$ noise is less important
than that of Ohmic fluctuations. However, in quantum dots
with large level spacings in the low-temperature and
low-field regime, these fluctuations could dominate over
the effect of  Ohmic fluctuations and eventually determine
the spin relaxation time.

\section{Conclusion}

We have shown how spin-orbit interaction in single electron
quantum dots induce pseudospin precession within Kramers'
doublets when the quantum dot is adiabatically shifted
along a path by electric fields. The precession depends
solely on the geometry of the path. We have analyzed the
resulting spin decoherence within a model restricted to two
orbital states and showed how random geometric phases
appear in this model. Then we have studied spin relaxation
and dephasing in a quantum dot considering all orbital
states. We derived an effective Hamiltonian for the lowest
Kramers' doublet. The geometrical spin precession induced
by at least two independent electric fields occurs around
non-commuting axis for different paths, so that it becomes
a strictly non-Abelian evolution. This has marked
consequences for the spin decoherence due to electric field
fluctuations. In particular the spin decay rate remains
finite as the external magnetic field vanishes. We
estimated the rates of relaxation processes due to the
coupling to phonons and to the ohmic environment.  We find
that, owing to the higher excitation density of ohmic
fluctuations at low frequencies as compared to phonon
fluctuations, the former dominate over the latter in the
low field relaxation rates and provide most likely the
leading relaxation channel. The underlying spin precession
mechanism could also be exploited to electrically
manipulate spins, namely if by applying electric fields the
electrons can be transported around $\sim\mathrm{\mu
m}$-sized loops.

{\bf Acknowledgements:} We like to thank W.A. Coish and D.
Loss for inspiring discussions. G.Z. was supported 
by the A. von Humboldt Foundation and by Hungarian Grants 
Nos. NF061726 and T046303.

\bibliographystyle{elsart-num}
\bibliography{Berry}

\end{document}